\RequirePackage{fix-cm}
\documentclass{svjour3}                     
\smartqed  
\usepackage{graphicx}
\usepackage{dcolumn}
\usepackage{bm}
\usepackage{amsmath}
\usepackage{amssymb}
\begin{document}

\title{Spin Correlations in $e^{+}e^{-}$ Pair Creation by Two-Photons and Entanglement  in QED}


\author{N. Yongram}


\institute{Department of Physics, Naresuan University, Phitsanulok 65000, Thailand\\
The Tah Poe Institute for Fundamental Study (TPTP-IF)\\
Naresuan University,  Phitsanulok 65000, Thailand and\\
Thailand Center of Excellence in Physics, CHE\\
 Ministry of Education, Bangkok 10400, Thailand\\
              \email{nattapongy@nu.ac.th}
}

\date{}

\maketitle

\begin{abstract}
Spin correlations of $e^{+}e^{-}$ pair productions of
two colliding photons are investigated and explicit expressions for
their corresponding probabilities are derived and found to be
\textit{energy} (speed) dependent, for initially \textit{linearly} and \textit{circularly polarized} photons,  different from those obtained by simply combining the spins of the relevant particles, for initially \textit{polarized} photons. These expressions also
depend on the angles of spin of $e^{+}$ (and/or of $e^{-}$), for initially {\it linearly polarized} photons, but not for {\it circularly polarized} photons, as a function of the energy.
It is remarkable that these explicit results obtained from quantum field theory show a clear violation of Bell's inequality of Local Hidden Variables theories at all {\it energies} beyond that of the threshold one for particle production, in support of quantum field theory in the relativistic regime. We hope that our explicit expression will lead to experiments, of the type described in the bulk of this paper, which can monitor energy (and speed) in polarization correlation experiments.
\end{abstract}

\section{INTRODUCTION}
\label{intro}
There have been many investigations over the years of particles's polarizations correlations \cite{EBM;1992,EBM;1994,EBM;1998,EBM_Ungkit;1994,McMaster_1961} in the light of Bell's inequality,
and Bell-like experiments have been proposed recently in high energy physics context \cite{Clauser_Horne;1974,Clauser_Shimoney;1978,Aspect;1982,Fry;1995,Kaday;1975,Osuch;1996,Irby;2003}. We have been interested in studying joint polarization
correlation probabilities of particles produced by initially \textit{polarized} as well as \textit{unpolarized} particles in fundamental processes of quantum electrodynamics (QED)
and the electro-weak theory \cite{NY_EBM;2003,EBM_NY;2004,EBM_NY;2005,NY;2008,NY_EBM_SS;2006}. Such studies, based on explicit computations in quantum field theory, show that the mere fact that particles emerging from a process have
non-zero speeds upon reaching the detectors, imply, in general, that their polarization correlations \textit{depend on speed} \cite{NY_EBM;2003,EBM_NY;2004,EBM_NY;2005,NY;2008,NY_EBM_SS;2006} and \textit{may also depend on the underlying
couplings} \cite{NY_EBM_SS;2006}. The explicit expressions of polarization correlations based on \textit{dynamical} computations following from field theory, are non-speculative,
involving no arbitrary input assumptions, and depend on speed, and possibly on the couplings as well. These are unlike formal arguments based simply on combining
the spins of the particles in question, which are of kinematical nature. In the limit of low energies, our earlier expressions \cite{NY_EBM;2003,EBM_NY;2004,EBM_NY;2005,NY;2008,NY_EBM_SS;2006} for the polarization correlations were shown to be reduced
to the simple method just mentioned by combining spins. In one of our previous investigations \cite{NY_EBM_SS;2006}, in which  $\mu^{+}\mu^{-}$ pair production in  $e^{+}e^{-}$ scattering was considered in the electro-weak theory. It was noted that, due to the threshold needed to create such a pair, the zero-energy limit may not be taken, and that the study of polarization correlations
by simply combining spins (without recourse to quantum field theory) has no meaning. The focus of this paper is the derivation of the explicit polarization correlation probabilities of
simultaneous measurements of spin of $e^{+}e^{-}$  pair production of two colliding photons, as well as the corresponding probabilities when only one of the  polarization of emerging $e^{+}$ (or $e^{-}$) is measured, in QED, for initially \textit{linearly polarized} as well as \textit{circularly polarized} photons, and particulary with initially \textit{unpolarized} photons, with emphasis put on the
\textit{energy} available in the process so that a detailed study can be carried out in the relativistic regime as well. Since the production of  $e^{+}e^{-}$ pair by the collision of two photons ($\gamma\gamma\to e^{+}e^{-}$) was originally proposed by Breit and Wheeler\,\cite{Breit_1934}, is so called ``\textit{Breit-Wheeler process}''. In the present paper, in the spirit of Bell type experiments, we are interested in joint {\it conditional} probability distributions of spin measurements rather than of cross sections \cite{Darbinian_1995}. Conditional in the sense that {\it given} such a process has occurred, {\it then} we ask as to what is the probability distribution of spin measurements correlations. It has still not being directly observed because of a relatively high energy threshold which is about $1.022$~MeV.  With the continuous effort of the experimental verification on Earth-based experiments, using  the technology based on free electron X-ray laser and its numerous applications, some first indications of its possible verification, have been reached. it also is one of most relevant elementary processes in high-energy astrophysics as well as in cosmology \cite{Nikishov_1961}. The QED calculations have been performed in stimulated Breit-Wheeler cross section \cite{Chen_1991,A. Hartin_2006} as well. The reasons for our present investigation are two fold. First several theoretical \cite{Motz_1964,Motz_1969,Page_1959} and experiments on $\gamma\gamma\to e^{+}e^{-}$ have been carried out over the years \cite{Maximon_1962}, and it is expected that our explicit new expression for the polarization correlations obtained, depending on energies, may lead to new experiments on spin correlations which monitor the energy (speed) of the underlying particles. Second, such a study may be relevant to experiments in the light of Bell's theorem (monitoring speed) as mentioned above and discussed below.

The relevant quantity of interest here in testing Bell's
inequality \cite{Clauser_Horne;1974,Clauser_Shimoney;1978} is, in a standard
notation,
\begin{align}
  S &= \frac{p_{12}(a_{1},a_{2})}{p_{12}(\infty,\infty)}
  -\frac{p_{12}(a_{1},a'_{2})}{p_{12}(\infty,\infty)}
  +\frac{p_{12}(a'_{1},a_{2})}{p_{12}(\infty,\infty)}
  +\frac{p_{12}(a'_{1},a'_{2})}{p_{12}(\infty,\infty)}
  \nonumber \\[0.5\baselineskip]
  &\quad
  -\frac{p_{12}(a'_{1},\infty)}{p_{12}(\infty,\infty)}
  -\frac{p_{12}(\infty,a_{2})}{p_{12}(\infty,\infty)}
  \label{Eqn1}
\end{align}
as is \emph{computed from} QED.   Here $a_{1}$,
$a_{2}$ \ $(a'_{1},a'_{2})$ specify directions along which the
polarizations of two particles are measured, with
$p_{12}(a_{1},a_{2})/p_{12}(\infty,\infty)$ denoting the joint
probability, and $p_{12}(a_{1},\infty)/p_{12}(\infty,\infty)$,\
$p_{12}(\infty,a_{2})/p_{12}(\infty,\infty)$ denoting the
probabilities when the polarization of only one of the particles
is measured.   [$p_{12}(\infty,\infty)$ is normalization factor.]
The corresponding probabilities as computed from QED
will be denoted by $P[\chi_{1},\chi_{2}]$, $P[\chi_{1},-]$,
$P[-,\chi_{2}]$ with $\chi_{1}$, $\chi_{2}$ denoting angles
specifying directions along which spin measurements are carried
out with respect to certain axes spelled out in the bulk of the
paper. To show that the QED process is in violation with
Bell's inequality of LHV, it is sufficient to find one set of
angles $\chi_{1}$, $\chi_{2}$, $\chi'_{1}$, $\chi'_{2}$, such that
$S$, as computed in QED, leads to a value of $S$
outside the interval $[-1,0]$. In this work, it is implicitly
assumed that the polarization parameters in the particle states
are directly observable and may be used for Bell-type measurements
as discussed. We show a clear violating of Bell's inequality for \textit{all} speeds in support of
quantum theory in the relativistic regime, i.e., of quantum field theory.

\begin{figure}
\includegraphics{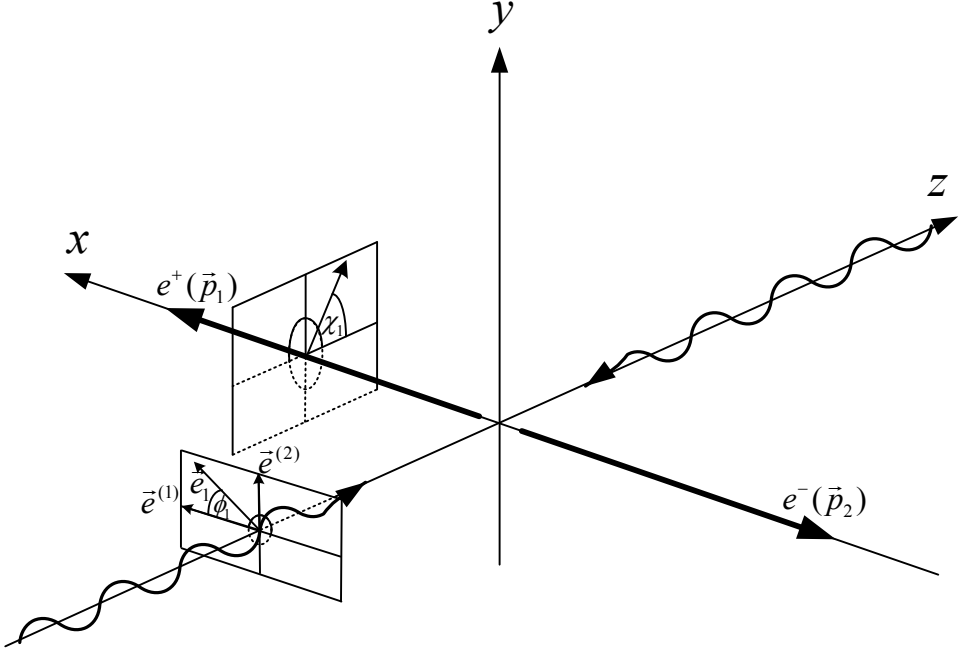}
\caption{\label{Fig1} The figure depicts $e^{+}e^{-}$ pair production by two-photons, with the initially photons
    moving along the $z$-axis, while the emerging electron and positron
   moving along the $x$-axis. The angle $\chi_{1}$ measured relative to
   the $z$-axis, denotes the orientation of spin of one of the emerging positron may make.  The unit vectors $\vec{e}^{\,(1)}$ and $\vec{e}^{\,(2)}$ are
   in $x$- and $y$-axis, respectively.}
\end{figure}

We consider the process of  $e^{+}e^{-}$ pair production of the collision of two photons, $\gamma(k_{1}) \gamma(k_{2})\rightarrow e^{+}(p_{1})e^{-}(p_{2})$, in the center of mass frame of the process (see Fig.~\ref{Fig1}), given by the amplitude of the process is well known\cite{Itzykson_1980}, up to an overall multiplicative factor irrelevant for the problem at hand.
\begin{eqnarray}\label{Eqn2}
\mathcal{A}\!\!\propto\!\overline{u}(p_{2})\!\!\left[\!\frac{\gamma^\mu\gamma
k_1\gamma^\nu}{2p_1k_1}\!+\!\frac{\gamma^\nu\gamma
k_2\gamma^\mu}{2p_1k_2}\!+\!\frac{\gamma^\mu
p^\nu_1}{p_1k_1}\!+\!\frac{\gamma^\nu p^\mu_1}{p_1k_2}\!\right]\!\!v(p_{1})
e^{\nu}_{1}e^{\mu}_{2}\;
\end{eqnarray}
where $e^\mu_1=(0,\vec{e}_{1})$, $e^\mu_{2}=(0,-\vec{e}_{2})$ are the
polarizations of two photons $\vec{k}_{1}$ and  $\vec{k}_{2}$,  respectively. It is convenient to compute the amplitude of the process above, be rewritten as $(j=1,2)$
 \begin{equation}\label{Eqn3}
\vec{e}_{j}\equiv(e^{(1)}_{j},e^{(2)}_{j},e^{(3)}_{j})
\end{equation}
and these polarizations will be specified later. For the four momenta of the initially photons and the emerging $e^{+}$ ($e^{-}$), respectively, given by
\begin{align}
\vec{k}_1&=\omega(0,0,1)=-\vec{k}_2\label{Eqn4}\\[0.5\baselineskip]
\vec{p}_{1}&=\omega\sqrt{1-\left(\frac{m_{e}}{\omega}\right)^{2}}(1,0,0)=-\vec{p}_{2}\label{Eqn5}
\end{align}
where $\omega$ is the energy of the photons, $m_{e}$ denote the mass of an electron and positron. The measurement of the spin projection of the positron is taken along
an axis making an angle $\chi_{1}$ with the $z$-axis and lying in parallel to the $x-z$ plane, as shown in Fig~\ref{Fig1},
\begin{equation}
  v(p_{1})=\sqrt{\frac{\omega}{m_{e}}}
  \begin{pmatrix}\rho\sigma_{1}\xi_{1}\\\\
  \xi_{1}
  \end{pmatrix} \;\text{and}\;
  u(p_{2})=\sqrt{\frac{\omega}{m_{e}}}
  \begin{pmatrix}\xi_{2}\\\\
  -\rho\sigma_{1}\xi_{2}
  \end{pmatrix}\label{Eqn6}
\end{equation}
where
\begin{equation}\label{Eqn7}
\rho=\frac{\sqrt{1-(m_{e}/\omega)}}{\sqrt{1+(m_{e}/\omega)}}
\end{equation}
and the direction of the spin of the electron makes an angle  $\chi_{2}$ with the $z$-axis. For the two-spinors, we have
\begin{align}
  \xi_{1}=
  \begin{pmatrix}-\mathrm{i}\cos\chi_{1}/2\\\\
  \sin\chi_{1}/2
  \end{pmatrix}\quad\text{and}\quad
  \xi_{2}=
  \begin{pmatrix}-\mathrm{i}\cos\chi_{2}/2\\\\
  \sin\chi_{2}/2
  \end{pmatrix}\label{Eqn8}
\end{align}
A straightforward computation gives the matrix elements as follows:
\begin{align}
  \overline{u}\left(\gamma^i\gamma^0\gamma^j\right)v&\sim
  2\rho\mathrm{i}\varepsilon_{ijk}[\xi^{\dag}_{2}\sigma_{k}\sigma_{1}\xi_{1}]\label{Eqn9}\\[0.5\baselineskip]
  \overline{u}\gamma^{i}v&\sim
  (1-\rho^2)[\xi^{\dag}_{2}\sigma_{1}\xi_{1}]\delta^{i1}+(1+\rho^2)[\xi^{\dag}_{2}\sigma_{2}\xi_{1}]\delta^{i2}\nonumber\\[0.5\baselineskip]
  &+(1+\rho^2)[\xi^{\dag}_{2}\sigma_{3}\xi_{1}]\delta^{i3}\label{Eqn10}\\[0.5\baselineskip]
  \overline{u}\left(\gamma^i\gamma^m\gamma^j\right)v&\sim\!\!
  \left(\!-\delta^{mj}\delta^{i1}\!\!\!-\!\delta^{mi}\delta^{j1}\!+\!
  \delta^{ji}\delta^{m1}\!\right)\!\!(1\!-\!\rho^2)[\xi^{\dag}_{2}\sigma_{1}\xi_{1}]\nonumber\\[0.5\baselineskip]
  &+\!\left(\!-\delta^{mj}\delta^{i2}\!\!-\!\delta^{mi}\delta^{j2}\!\!+\!
  \delta^{ji}\delta^{m2}\right)\!\!(1\!+\!\rho^2)\![\xi^{\dag}_{2}\sigma_{2}\xi_{1}]\nonumber\\[0.5\baselineskip]
  &+\!\left(\!-\delta^{mj}\delta^{i3}\!\!-\!\delta^{mi}\delta^{j3}\!\!+\!
  \delta^{ji}\delta^{m3}\right)\!(1\!+\!\rho^2)\![\xi^{\dag}_{2}\sigma_{3}\xi_{1}]\nonumber\\[0.5\baselineskip]
  &-\mathrm{i}(1-\rho^2)\varepsilon_{mji}[\xi^{\dag}_{2}\xi_{1}]\label{Eqn11}
\end{align}
where $k=1,2,3$. The matrix element above and Eq.~(\ref{Eqn3}) give the amplitude $\mathcal{A}$ in Eq.~(\ref{Eqn2}) as
\begin{align}\label{Eqn12}
\mathcal{A}&\propto-\mathrm{i}(1-\rho^2)\vec{k}\cdot(\vec{e}_1\times\vec{e}_2)[\xi^{\dag}_{2}\xi_{1}]\nonumber\\[0.5\baselineskip]
&+(1-\rho^2)\sqrt{1-\left(\frac{m_{e}}{\omega}\right)^{2}}\left(e^{(1)}_1e^{(1)}_2+e^{(1)}_{2}e^{(1)}_1\right)[\xi^{\dag}_{2}\sigma_{1}\xi_{1}]
\nonumber\\[0.5\baselineskip]
&+(1+\rho^2)\sqrt{1-\left(\frac{m_{e}}{\omega}\right)^{2}}\left(e^{(2)}_1e^{(1)}_2+e^{(2)}_{2}e^{(1)}_1\right)[\xi^{\dag}_{2}\sigma_{2}\xi_{1}]
\end{align}
which depends on the specification of the polarization of the initial photons in the process.

In Sec.~\ref{Polarization Correlations}, the amplitude in Eq.~(\ref{Eqn12}) is applied to compute exact joint probabilities of simultaneous measurement of spins
of $e^{+}e^{-}$ pair productions of two colliding linear and circular polarized, as well as unpolarized photons. It is also applied to compute of the corresponding probabilities of
one of spin of $e^{+}$ (or $e^{-}$). We then simulated our computation to show violation of Bell's inequality of LHV theories, which may lead to new experiments on spin correlations which monitor the energy (speed) of the underlying particles in Sec.~\ref{Simulations and Discussions}. Finally, we conclude our results in Sec.~\ref{Conclusions}.
\section{POLARIZATION CORRELATIONS IN BREIT-WHEELER PROCESS}
\label{Polarization Correlations}
\subsection{initially linearly polarized photons}
\label{linear}
Now we consider the process, $\gamma\gamma\to e^{+}e^{-}$,  with linear polarization of initial photons. The photon polarization vector $\vec{e}_{j}$
is given by  ($j=1,2$)
\begin{equation}\label{Eqn13}
\vec{e}_{j}=\cos\phi_{j}\vec{e}^{\,(1)}+\sin\phi_{j}\vec{e}^{\,(2)}
\end{equation}
where $\vec{e}^{\,(1)}$ and $\vec{e}^{\,(2)}$ are unit vectors in the plane of interaction and perpendicular to
it, respectively. These photon linear polarization vectors are given by
\begin{align}\label{Eqn14}
\vec{e}^{\,(1)}=(1,0,0)\;\text{and}\;\vec{e}^{\,(2)}=(0,1,0)
\end{align}
and $\phi$ is the angle between $\vec{e}$ and the $\vec{e}^{\,(1)}$ axis, as shown in Fig.~\ref{Fig1}. In this case,
the planes of photon polarization at an angle of $\phi_{1}$ and $\phi_{2}$  are equal to $\pi/4$ with $x$-axis, then Eq.~(\ref{Eqn13}) gives
the polarization of the initial photons as $\vec{e}_{1}=\frac{1}{\sqrt{2}}(1,1,0)=-\vec{e}_{2}$. By substituting
the polarization state of initial photons in Eq.~(\ref{Eqn12}), we have the amplitude of the process of the initial linear polarized photons, given by
\begin{equation}\label{Eqn15}
\mathcal{A}_{\text{linear}}\propto\frac{\mathrm{i}m_{e}}{\omega\left[1+\left(\frac{m_{e}}{\omega}\right)\right]}\sin\left(\frac{\chi_1-\chi_2}{2}\right)
-\frac{1}{\left[1+\left(\frac{m_{e}}{\omega}\right)\right]}\sin\left(\frac{\chi_1+\chi_2}{2}\right)
\end{equation}
Using notation $F[\chi_{1},\chi_{2}]$ for the absolute value square of the right-hand side of Eq.~(\ref{Eqn15}),
the conditional joint probability distribution of spin measurements along the directions specified by angle
$\chi_{1}$, $\chi_{2}$ is given by
 \begin{equation}
P[\chi_{1},\chi_{2}]=\frac{F[\chi_{1},\chi_{2}]}{N(\omega)} \label{Eqn16}
\end{equation}
The normalization factor $N(\omega)$ is obtained by summing over all the polarizations of the emerging particles. This is
equivalent to summing of $F[\chi_{1},\chi_{2}]$ over the pairs of angles
 \begin{equation}
(\chi_{1},\chi_{2}),\quad(\chi_{1}+\pi,\chi_{2}),\quad(\chi_{1},\chi_{2}+\pi),\quad(\chi_{1}+\pi,\chi_{2}+\pi) \label{Eqn17}
\end{equation}
This leads to
\begin{align}
N(\omega)&=F[\chi_{1},\chi_{2}]+F[\chi_{1}+\pi,\chi_{2}]+F[\chi_{1},\chi_{2}+\pi]+P[\chi_{1}+\pi,\chi_{2}+\pi]\nonumber\\[0.5\baselineskip]
&=\frac{2\left[1+\left(\frac{m_{e}}{\omega}\right)^{2}\right]}{\left[1+\left(\frac{m_{e}}{\omega}\right)\right]^{2}} \label{Eqn18}
\end{align}
giving
\begin{equation}
P[\chi_1,\chi_2]=\frac{m^{2}_{e}}{2\omega^{2}\left[1+\left(\frac{m_{e}}{\omega}\right)^{2}\right]}\sin^{2}\left(\frac{\chi_{1}-\chi_{2}}{2}\right)
+\frac{1}{2\left[1+\left(\frac{m_{e}}{\omega}\right)^{2}\right]}\sin^{2}\left(\frac{\chi_{1}+\chi_{2}}{2}\right)
\label{Eqn19}
\end{equation}
If only one of the spins is measured corresponding to $\chi_{1}$, the probability can be written as
\begin{align}
P[\chi_{1},-]&= P[\chi_{1},\chi_{2}]+P[\chi_1,\chi_{2}+\pi]\nonumber\\[0.5\baselineskip]
&= \frac{1}{2}\label{Eqn20}
\end{align}
Similarly, if only one of the spins is measured corresponding to $\chi_{2}$, the probability can be written as
\begin{align}
P[-,\chi_{2}]&= P[\chi_{1},\chi_{2}]+P[\chi_{1}+\pi,\chi_{2}]\nonumber\\[0.5\baselineskip]
&= \frac{1}{2}\label{Eqn21}
\end{align}

\subsection{\label{circularly}initially circularly polarized photons}
Here we turn to consider the production of two initially photons $\vec{k}_{1}$ and  $\vec{k}_{2}$ in the process,  with the right-handed and left-handed circular polarization,
are specified by the vectors
\begin{align}\label{Eqn22}
\vec{e}_{1}=\frac{1}{\sqrt{2}}(1,\mathrm{i},0)\;\text{and}\;\vec{e}_{2}=\frac{1}{\sqrt{2}}(1,-\mathrm{i},0)
\end{align}
respectively, producing $e^{+}e^{-}$ pairs, and place detectors for the latter at opposite end of the $x$-axis, as in Fig.~\ref{Fig1}, this gives
\begin{equation}\label{Eqn23}
\mathcal{A}_{\text{cir}}\propto\omega\cos\left(\frac{\chi_1-\chi_2}{2}\right)
+\mathrm{i}\sqrt{1-\left(\frac{m_{e}}{\omega}\right)^{2}}\sin\left(\frac{\chi_1-\chi_2}{2}\right)
\end{equation}
By using notations in Eqs.~(\ref{Eqn16})--(\ref{Eqn18}) to compute the joint probability of spin correlations of $e^{+}e^{-}$ produced by two circularly polarized photons, this gives
\begin{align}
P[\chi_1,\chi_2]&=\frac{\omega^{2}+\left[(1-\omega^{2})-\left(\frac{m_{e}}{\omega}\right)^{2}\right]\sin^{2}\left(\frac{\chi_1-\chi_2}{2}\right)}{2\left[2\omega^{2}+\left[(1-\omega^{2})-\left(\frac{m_{e}}{\omega}\right)^{2}\right]\right]}
 \label{Eqn24}\\[0.5\baselineskip]
P[\chi_1,-]&= \frac{1}{2}\label{Eqn25}\\[0.5\baselineskip]
P[-,\chi_2]&= \frac{1}{2}\label{Eqn26}
\end{align}

\subsection{\label{unpolarized}initially unpolarized photons}
Finally, we consider the process $\gamma\gamma \rightarrow e^{+}e^{-}$, in the
c.m., with initially unpolarized photons with some over all polarization of the photons
\begin{equation}
\sum\limits_{\text{pol}}{e^{i}_{2}e^{j}_{1}}=\delta^{ij}-n^{i}_{2}n^{j}_{1}\label{Eqn27}
\end{equation}
where $\vec{n}=\vec{k}/|\vec{k}|$ and $i,j=1,2$, using the identity $\sum\limits_{\text{pol}}{\vec{k}\cdot(\vec{e}_{1}\times\vec{e}_{2})}=0$, this gives
\begin{equation}\label{Eqn28}
\mathcal{A}_{\text{unpol}}\propto\xi^{\dag}_{2}\xi_{1}\left\{\begin{pmatrix}1\\0\end{pmatrix}_{2}\begin{pmatrix}0&1\end{pmatrix}_{1}
+\begin{pmatrix}0\\1\end{pmatrix}_{2}\begin{pmatrix}1&0\end{pmatrix}_{1}\right\}
\end{equation}
(see Fig.~\ref{Fig1}), generating an energy independent (normalized) entangled state for $e^{+}e^{-}$ given by
\begin{equation}\label{Eqn29}
|\psi\rangle=\frac{1}{\sqrt{2}}\left[\begin{pmatrix}1\\0\end{pmatrix}_{2}\begin{pmatrix}0&1\end{pmatrix}_{1}
+\begin{pmatrix}0\\1\end{pmatrix}_{2}\begin{pmatrix}1&0\end{pmatrix}_{1}\right]
\end{equation}
Therefore the joint probability of  spin correlations of $e^{+}e^{-}$ is given by
\begin{align}
P[\chi_1,\chi_2]&=\|\xi^{\dag}_{2}\xi_{1}|\psi\rangle\|^{2}\nonumber\\[0.5\baselineskip]
&=\frac{1}{2}\sin^{2}\left(\frac{\chi_1-\chi_2}{2}\right)\label{Eqn30}
\end{align}
and
\begin{align}
P[\chi_{1},-]&= \|\xi_{1}|\psi\rangle\|^{2}=\frac{1}{2}\label{Eqn31}\\[0.5\baselineskip]
P[-,\chi_{2}]&= \|\xi^{\dag}_{2}|\psi\rangle\|^{2}=\frac{1}{2}\label{Eqn32}
\end{align}
$P[\chi_{1},-]$ is also \textit{equivalently} obtained by summing $P[\chi_{1},\chi_{2}]$ over
\begin{equation}\label{Eqn33}
\chi_{1},\quad\chi_{2}+\pi
\end{equation}
for any arbitrarily chosen $\chi_{2}$, i.e., as in Eq.~(\ref{Eqn20}) and similarly for  $P[-,\chi_{2}]$.
As the result of our calculation, we found that the joint polarization correlation probabilities of the initial unpolarized photons
are independent of energy, as in the combining spin of the kinematic considerable.
\section{SIMULATIONS AND DISCUSSIONS}\label{Simulations and Discussions}
In this section, we simulate the joint polarization correlation probabilities of $e^{+}e^{-}$ pair productions of two colliding photons,
for initially linearly and circularly polarized photons, computed in the previous section to show a clear violation
of the relevant Bell-like inequality as a function of the energy of initially photons.
We note the important statical property that
 \begin{equation}\label{Eqn34}
P[\chi_{1},\chi_{2}]\neq P[\chi_{1},-]P[-,\chi_{2}]
\end{equation}
in general, showing obvious correlations occurring between the two spins. The indicator $S$ in Eq.~(\ref{Eqn1}) computed according to the
probabilities of simultaneous measurement of the spins of $e^{+}e^{-}$ $P[\chi_{1},\chi_{2}]$ in Eqs.~(\ref{Eqn19}), (\ref{Eqn24}), (\ref{Eqn30}), and
the corresponding probabilities when only one of the spins of $e^{+}$ (or $e^{-}$) $P[\chi_{1},-]$ in Eqs.~(\ref{Eqn20}), (\ref{Eqn25}), (\ref{Eqn31}),
as well as $P[-,\chi_{2}]$ in Eqs.~(\ref{Eqn21}),  (\ref{Eqn26}),  (\ref{Eqn32}), may be readily evaluated. To show violation of Bell's inequality of LHV theories, it is
sufficient to find four angles $\chi_{1}$, $\chi_{2}$, $\chi'_{1}$,
$\chi'_{2}$ at accessible energies, for which $S$ falls outside the
interval $[{-}1,0]$. We perform such a simulation fitting, presenting our results, and discussing their physical implications.

\begin{figure}
\centering
\includegraphics[width=0.9\textwidth]{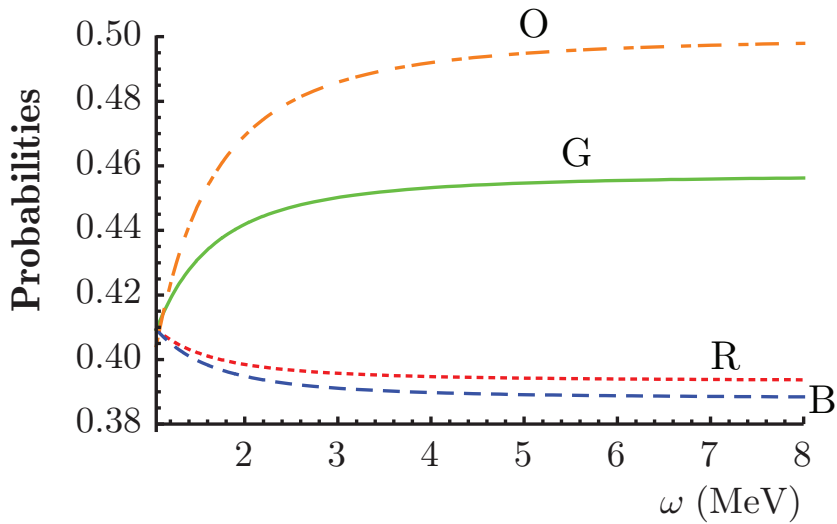}
\caption{\label{Fig2} The probabilities of simultaneous measurement of spin correlations of $e^{+}e^{-}$ pair productions
of two-colliding linearly polarized photons, $P[\chi'_{1},\chi'_{2}]$, as a function of the energies of photons. The dashed-dotted line (O) is for the set of angles $\chi'_{1}=67^{\circ}$,  $\chi'_{2}=213^{\circ}$. The solid line (G) is for the set of angles $\chi'_{1}=90^{\circ}$, $\chi'_{2}=270^{\circ}$. The dotted line (R) is for the set of angles $\chi'_{1}=30^{\circ}$, $\chi'_{2}=140^{\circ}$. The dashed line (B) is for the set of angles $\chi'_{1}=30^{\circ}$, $\chi'_{2}=153.5^{\circ}$.}
\end{figure}

As shown in Table~\ref{tab1}, we found that the deriving probabilities in Sec.~\ref{linear}, for initially linearly polarized photons, give $S$ outside the interval $[{-}1,0]$ from below for four angles $\chi_{1}$, $\chi_{2}$, $\chi'_{1}$, $\chi'_{2}$ at accessible energies. The plotting in Fig.~\ref{Fig2} show that the probability $P[\chi'_{1},\chi'_{2}]$
in Eq.~(\ref{Eqn19}) is a varying function of the energy of initial photons. These probabilities, for the set of angles ($180^{\circ}<\chi'_{1}<360^{\circ}$, $0^{\circ}\leq\chi'_{2}\leq360^{\circ}$) and ($0^{\circ}\leq\chi'_{1}\leq360^{\circ}$, $180^{\circ}<\chi'_{2}<360^{\circ}$), are rapidly increasing, $\omega\leq$3.5  MeV,  after that they are slightly increasing with energy of the initially photons. On the other hand, for the set of angles ($0^{\circ}<\chi'_{1}<180^{\circ}$, $0^{\circ}\leq\chi'_{2}\leq360^{\circ}$) and ($0^{\circ}\leq\chi'_{1}\leq360^{\circ}$, $0^{\circ}<\chi'_{2}<180^{\circ}$), they are slightly decreasing with energy of the initially photons.  However, we found that $S$ are varying functions of the set of angles $\chi'_{1}$,
$\chi'_{2}$ when angles $\chi_{1}=0^{\circ}$, $\chi_{2}=45^{\circ}$ and the energy are fixed which can see are decreasing functions of the energy
of  the initially photons when the angles are fixed (as shown in Fig.~\ref{Fig4}).  Such the results as  the probabilities and the indicator $S$  depend on the angles of spin of $e^{+}e^{-}$
and the energy of initially photons in processes.
\begin{table}
\caption{\label{tab1}The indicator $S$ computed according to the probabilities of spin correlations of $e^{+}e^{-}$ pair productions
of two-colliding linearly polarized photons are simulated. For example, the four angles $\chi_{1}=0^{\circ}$, and $\chi_{2}$, $\chi'_{1}$,
$\chi'_{2}$ are shown in a following table. R, B, G and O denotes  Red, Blue, Green and Orange, respectively.}
\begin{center}
 \begin{tabular}{cccccc}
\hline
line& $\omega$(MeV) &$\chi_{2}$(degree) &$\chi^{\prime}_{1}$(degree)&$\chi^{\prime}_{2}$(degree)& $S$ \\
\hline
-&1.05 & 45 & 15&180& -1.37576\\
R&1.05 & 45 & 30& 140& -1.36279\\
B&1.05 & 45 & 30& 153.5& -1.35814\\
G&1.05 & 45 & 67& 213& -1.30592\\
O&1.05 & 45 & 90& 270& -1.03585\\
-&5.00 & 45 & 15&180& -1.39234\\
-&10.00 & 45 & 15&180& -1.39304\\
-&35.00 & 45 & 15&180&-1.39326\\
-&$46.60\times10^{3}$ & 45 & 15&180&-1.39328\\
\hline
\end{tabular}
\end{center}
\end{table}

\begin{figure}
\centering
\includegraphics[width=0.8\textwidth]{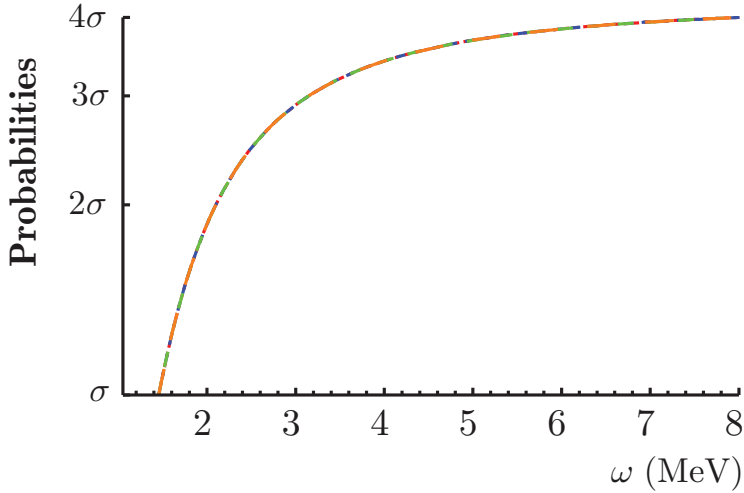}
\caption{\label{Fig3} The probabilities of simultaneous measurement of spin correlations of $e^{+}e^{-}$ pair productions
of two-colliding circularly polarized photons, $P[\chi'_{1},\chi'_{2}]$, as a function of the energies of photons. We have taken the angles as in Table~\ref{tab2}.}
\end{figure}

\begin{figure}
\centering
\includegraphics[width=0.9\textwidth]{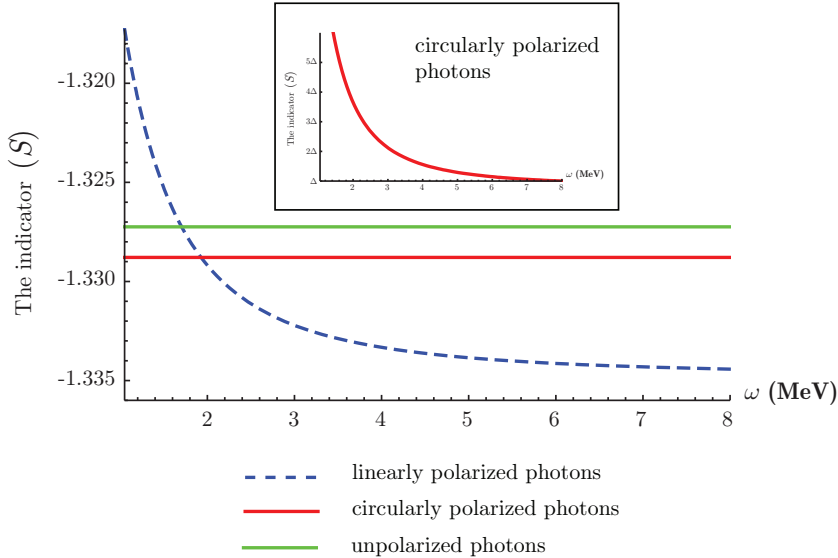}
\caption{\label{Fig4} The plotting  of the indicator $S$  as a functions of the energy of initially photons, for four set of angles $\chi_{1}=0^{\circ}$, $\chi_{2}=45^{\circ}$, $\chi'_{1}=15^{\circ}$, $\chi'_{2}=140^{\circ}$ of spin of $e^{+}e^{-}$ for initially linearly polarized photon, as well as for four set of angles $\chi_{1}=140^{\circ}$, $\chi_{2}=155^{\circ}$, $\chi'_{1}=15^{\circ}$,
$\chi'_{2}=50^{\circ}$ of spin of $e^{+}e^{-}$ for initially circularly polarized photon. In particular for four set of angles $\chi_{1}=140^{\circ}$, $\chi_{2}=23^{\circ}$, $\chi'_{1}=67^{\circ}$,
$\chi'_{2}=132^{\circ}$ of spin of $e^{+}e^{-}$ for initially unpolarized photon.}
\end{figure}

As shown in Table~\ref{tab2} we found that the deriving probabilities in Sec.~\ref{circularly}, for initially circularly polarized photons, gives $S$ outside the interval $[{-}1,0]$ from below for four angles $\chi_{1}$, $\chi_{2}$, $\chi'_{1}$,
$\chi'_{2}$ at accessible energies.  The plotting in Fig.~\ref{Fig3} show that  the probability $P[\chi'_{1},\chi'_{2}]$
in Eq.~(\ref{Eqn24}) is a varying function of the energy of initially photons. These probabilities, for four set of angles $\chi'_{1}$ and $\chi'_{2}$ (see in Table~\ref{tab2}), are rapidly increasing , $\omega\leq$5 MeV, after that they are slightly increasing with the energy of the initially photons. These plotting lies on the same line, even though  the set of angles $\chi'_{1}$ and $\chi'_{2}$ were changed. We also found that $S$ are varying functions of angles $\chi'_{1}$, $\chi'_{2}$ when angles $\chi_{1}=0^{\circ}$, $\chi_{2}=155^{\circ}$ and the energy are fixed. These $S$ are a rapidly decreasing with the increasing energy of initially photons when angles $\chi_{1}=0^{\circ}$, $\chi_{2}=155^{\circ}$, $\chi'_{1}=15^{\circ}$ and $\chi'_{2}=50^{\circ}$ are fixed as shown in Table~\ref{tab3}, and plotted in Fig.~\ref{Fig4}.
Such the results as the indicator $S$  depend on the angles of the measurement  of spin of $e^{+}e^{-}$
and the energy of initially photons, but the probabilities only depend on the energy of initially photons in processes.
\begin{table}
\caption{\label{tab2}The indicator $S$ computed according to the probabilities of spin correlations of $e^{+}e^{-}$ pair productions
of two-colliding circularly polarized photons are simulated. For example, the four angles $\chi_{1}=0^{\circ}$, and $\chi_{2}$, $\chi'_{1}$,
$\chi'_{2}$ are shown in a following table. R, B, G and O denotes  Red, Blue, Green and Orange, respectively.}
\begin{center}
\begin{tabular}{cccccc}
\hline
line &$\omega$(MeV) &$\chi_{2}$(degree) &$\chi^{\prime}_{1}$(degree)&$\chi^{\prime}_{2}$(degree)& $S$ \\
\hline
R&1.05 & 155 & 15& 50& -1.32878\\
G&1.05 & 155 & 45& 10& -1.30828\\
B&1.05 & 155 & 85& 50& -1.05177\\
O&1.05 & 155 & 90& 55& -1.01432\\
\hline
\end{tabular}
\end{center}
\end{table}
\begin{table}
\caption{\label{tab3}The indicator $S$ computed according to the probabilities of spin correlations of $e^{+}e^{-}$ pair productions
of two-colliding circularly polarized photons are simulated. For example, the four angles $\chi_{1}=0^{\circ}$, and $\chi_{2}=155^{\circ}$, $\chi'_{1}=15^{\circ}$,
$\chi'_{2}=50^{\circ}$ and varying photon energies are shown in a following table.}
\begin{center}
\begin{tabular}{cc}
\hline
$\omega$(MeV) &  $S$ \\
\hline
1.05 & -1.328 784 959 959 7406\\
5.00  & -1.328 784 959 960 4962\\
10.00 & -1.328 784 959 960 5301\\
35.00 & -1.328 784 959 960 5410\\
$46.60\times10^{3}$ &-1.328 784 959 960 5420\\
\hline
\end{tabular}
\end{center}
\end{table}
\begin{table}
\caption{\label{tab4}The value of the probabilities of spin correlations of $e^{+}e^{-}$ pair productions
of two-colliding circularly polarized photons are shown in Fig.~\ref{Fig3}.}
\begin{center}
\begin{tabular}{cc}
\hline
 variables & Probabilities\\
\hline
$\sigma$ &   0.454 788 011 072 0782\\
$2\sigma$ & 0.454 788 011 072 1606\\
$3\sigma$ & 0.454 788 011 072 2077\\
$4\sigma$ & 0.454 788 011 072 2416\\
\hline
\end{tabular}
\end{center}
\end{table}

In Table~\ref{tab4} we found that the deriving probabilities in Sec.~\ref{unpolarized}, for initially unpolarized photons, also give $S$ outside the $[{-}1,0]$ from below for four angles $\chi_{1}$, $\chi_{2}$, $\chi'_{1}$,
$\chi'_{2}$, and  {\it independent on energies} (shown in Table~\ref{tab5}). These $S$ are plotted in Fig.~\ref{Fig4} as well.  The indicator $S$ and the probability $P[\chi_{1},\chi_{2}]=(1/2)\sin^{2}[(\chi_{1}-\chi_{2})/2]$  are {\it independent of the energy} of initially photons, but they only depend on
the angles $\chi_{1}$, $\chi_{2}$ of the measurement of   $e^{+}e^{-}$ spins respectively.
\begin{table}
\caption{\label{tab5}The indicator $S$ computed according to the probabilities of spin correlations of $e^{+}e^{-}$ pair productions
of two-colliding unpolarized photons are simulated. For example, the four angles $\chi_{1}=0^{\circ}$, and $\chi_{2}$, $\chi'_{1}$,
$\chi'_{2}$ are shown in a following table.}
\begin{center}
\begin{tabular}{ccccc}
\hline
 $\omega$(MeV) &$\chi_{2}$(degree) &$\chi^{\prime}_{1}$(degree)&$\chi^{\prime}_{2}$(degree)& $S$ \\
\hline
{\it independent} & 85 & 25& 181& -1.14675\\
{\it independent} & 67 & 55& 181& -1.34218\\
{\it independent} & 23 & 45& 180 & -1.46192\\
\hline
\end{tabular}
\end{center}
\end{table}
Finally,  we have plotted the indicator $S$  as a functions of the energy of initially photons, for four set of angles $\chi_{1}$, $\chi_{2}$, $\chi'_{1}$,
$\chi'_{2}$, as shown in Fig.~\ref{Fig4}. Such the result as the indicator $S$ for initially \textit{linearly polarized} photons is faster decreasing with the energy than  that of initially \textit{circularly polarized} photons  when set of angles is fixed.
\begin{table}
\caption{\label{tab6}The value of the probabilities of spin correlations of $e^{+}e^{-}$ pair productions
of two-colliding circularly polarized photons are shown in Fig.~\ref{Fig4}.}
\begin{center}
\begin{tabular}{cc}
\hline
 variables & The indicator ({\it S}) \\
\hline
$\triangle$ &   -1.328 784 959 960 5240\\
$2\triangle$ & -1.328 784 959 960 4287\\
$3\triangle$ & -1.328 784 959 960 3334\\
$4\triangle$ & -1.328 784 959 960 2382\\
$5\triangle$ & -1.328 784 959 960 1430\\
\hline
\end{tabular}
\end{center}
\end{table}

\section{CONCLUSIONS}\label{Conclusions}
We have investigated and derived, in detail, the explicit polarization correlation probabilities of simultaneous measurement of spins of $e^{+}e^{-}$ pair productions by two colliding photons  for initially \textit{linearly} and \textit{circularly polarized} photons, emphasizing their dependence on energy for initially {\it polarized} photons. The expressions were obtained in the (relativistic) QED setting. The necessity of such a study within the realm of quantum field theory cannot be overemphasized, as estimates of such correlations from simply combining spins (as is often done), have no meaning, as they do not involve dynamical considerations. The relevant dynamics is, of course, dictated directly from quantum field theory. The explicit expression for the polarization correlation obtained is interesting in its own right, and they may also lead to experiments that investigate such correlations by monitoring energy (speed), for initially {\it linearly} and {\it circularly polarized} photons, but not for {\it unpolarized} ones. The simulation study carried out in Sec.~\ref{Simulations and Discussions}, shows that these expressions also depend on the angles of spin of $e^{+}$ (and/or $e^{-}$), for initially {\it linearly polarized} photons, but not for {\it circularly polarized} photons, as a function of the energy. Our results may also be relevant in the realm of Bell's inequality with emphasis on relativistic aspects of quantum theory, that is, of quantum field theory. Our expressions have shown clear violations of Bell's inequality of LHV theories, in support of quantum theory in the relativistic regime. In recent years, several experiments have been already performed  (cf.
\cite{Aspect;1982,Fry;1995,Kaday;1975,Osuch;1996,Irby;2003}) on particles' polarization correlations. It is expected that the novel properties recorded here by explicit calculations following directly from field theory (which is based on the principle of relativity and quantum theory) will lead to new experiments on polarization correlations for monitoring speed in the light of Bell's theorem. We hope that these computations, within the general setting of quantum field theory, will also be useful in areas of physics such as quantum teleportation and quantum information.
\section*{ACKNOWLEDGMENTS}
I would like to thank Prof. Dr. E.~B.~ Manoukian, Dr.~Burin~Gumjudpai and Dr.~ Ong-on Topon for discussions, guidance and for carefully reading the manuscript. I also thanks Mr. Chakkrit Kaeonikhom for correcting graph. I would like to acknowledge the Thailand Research Fund for a New Researchers Grant (MRG5080288), the Thailand Center of Excellence in Physics and the Faculty of Science of Naresuan University for supports.

\end{document}